\documentclass[pdftex,twocolumn,epjc3]{svjour3}          

\RequirePackage[T1]{fontenc}

\smartqed  
\RequirePackage{graphicx}
\RequirePackage{mathptmx}      
\RequirePackage{flushend}
\RequirePackage[numbers,sort&compress]{natbib}
\RequirePackage[colorlinks,citecolor=blue,urlcolor=blue,linkcolor=blue]{hyperref}

\journalname{Eur. Phys. J. C}

\begin{document}

\title{Cosmological perturbations and  dynamical analysis for interacting quintessence}


\author{Ricardo G. Landim\thanksref{e1,addr1}
}

\thankstext{e1}{ricardo.landim@tum.de}

\institute{Physik Department T70, James-Franck-Strasse,
Technische Universit\"at M\"unchen, 85748 Garching, Germany\label{addr1}
}

\date{Received: date / Accepted: date}

\maketitle

\begin{abstract}
We present the dynamical analysis for interacting quintessence, considering linear cosmological perturbations. Matter perturbations improve the background analysis  and viable critical points describing the transition of the three cosmological eras are found.  The stability of those  fixed points are similar to previous studies in the literature, for both coupled and uncoupled cases, leading to a late-time attractor.  \end{abstract}

\section{Introduction}

Observations of Type IA Supernova indicate that the Universe undergoes an accelerated expansion \cite{reiss1998, perlmutter1999}, which is dominant today ($\sim$ 68\%) \cite{Aghanim:2018eyx}. Ordinary matter represents only $5\%$ of the energy content of the Universe, and the remaining $27\%$ is the still unknown  dark matter (DM). The nature of the dark sector is one of the biggest challenges in the modern cosmology, whose plethora of dark energy (DE) candidates  include  scalar fields
\cite{peebles1988,ratra1988,Frieman1992,Frieman1995,Caldwell:1997ii,Padmanabhan:2002cp,Bagla:2002yn,ArmendarizPicon:2000dh,Brax1999,Copeland2000,Vagnozzi:2018jhn},
vector fields
\cite{Koivisto:2008xf,Bamba:2008ja,Emelyanov:2011ze,Emelyanov:2011wn,Emelyanov:2011kn,Kouwn:2015cdw,Landim:2016dxh},
holographic dark energy
\cite{Hsu:2004ri,Li:2004rb,Pavon:2005yx,Wang:2005jx,Wang:2005pk,Wang:2005ph,Wang:2007ak,Landim:2015hqa,Li:2009bn,Li:2009zs,Li:2011sd,Saridakis:2017rdo,Mamon:2017crm,Mukherjee:2016lor,Feng:2016djj,Herrera:2016uci,Forte:2016ben},
models of false vacuum decay
\cite{Szydlowski:2017wlv,Stachowski:2016zpq,Stojkovic:2007dw,Greenwood:2008qp,Abdalla:2012ug,Shafieloo:2016bpk,Landim:2016isc},
modifications of gravity and different kinds of cosmological fluids
\cite{copeland2006dynamics, dvali2000, yin2005}.  In addition, the two
components of the dark sector may interact with each other
\cite{Wetterich:1994bg,Amendola:1999er,Guo:2004vg,Cai:2004dk,Guo:2004xx,Bi:2004ns,Gumjudpai:2005ry,yin2005,Wang:2005jx,Wang:2005pk,Wang:2005ph,Wang:2007ak,Costa:2013sva,Abdalla:2014cla,Costa:2014pba,Costa:2016tpb,Marcondes:2016reb,Landim:2016gpz,Wang:2016lxa,Farrar:2003uw,Abdalla:2012ug,micheletti2009,Yang:2017yme,Marttens:2016cba,Yang:2017zjs,Costa:2018aoy,Yang:2018euj},
since their densities are comparable and the interaction can eventually
alleviate the coincidence problem \cite{Zimdahl:2001ar,Chimento:2003iea}.  

When a scalar field is in the presence of a barotropic fluid the relevant evolution equations can be converted   into an autonomous system and the asymptotic states of the cosmological models can be analysed. Such approach is well-known, at the background level, 
for uncoupled dark energy (quintessence, tachyon field and phantom field  for instance \cite{copeland1998,ng2001,Copeland:2004hq,Zhai2005,DeSantiago:2012nk}) and coupled dark energy \cite{Amendola:1999er,Gumjudpai:2005ry,TsujikawaGeneral,amendola2006challenges,ChenPhantom,Mahata:2015lja,Landim:2015poa,Landim:2015uda,Landim:2016dxh,Landim:2016gpz}. On the other hand, cosmological perturbations were only studied using dynamical analysis for $\Lambda$CDM \cite{Alho:2019jho,Basilakos:2019dof} and quintessence \cite{Basilakos:2019dof}. The role of cosmological perturbations  in uncoupled and coupled quintessence (with diverse forms of interactions between the dark sector) has been investigated in several works \cite{Bertolami:1999dp,Baccigalupi:2001aa,Dave:2002mn, Pettorino:2004zt,Brookfield:2005bz,Koivisto:2005nr,Lee:2006za,Olivares:2006jr,Pettorino:2008ez,Tarrant:2011qe,Sefusatti:2011cm,Liu:2019ygl}, whose aim is to constrain the free parameters of the model using  sets of observations. Results from dynamical analysis are also usually employed in such works, as in \cite{LopesFranca:2002ek}, for instance. Therefore, it is interesting to improve the background analysis in order to understand whether the fixed points are viable to describe each one of the cosmological eras of the Universe or not. This can be done  if one takes cosmological perturbations into account.  In this paper we go in this direction, analysing interacting quintessence with cosmological perturbations, in the light of dynamical systems theory. 
Our findings mostly agree with previous results in the literature, where it was used only background equations, for coupled (and uncoupled) quintessence, including the stability of the fixed points. One of those critical points, however, no longer can describe a DE-dominated Universe, when one uses cosmological perturbations.

The rest of the paper is organized as follows. In Sect. \ref{de} we present the basics of the interacting DE and the dynamical analysis theory. In Sect.  \ref{quintphantom} we present the dynamics of the canonical scalar field, with the correspondent equations for the background and for linear perturbations. We use the dynamical system theory in Sect. \ref{auto} to study interacting quintessence at cosmological perturbation level, analysing the critical points and their stabilities. Section \ref{conclu} is reserved for conclusions. We use Planck units ($\hbar=c=M_{pl}=1$) throughout the text.

\section{Interacting dark energy and dynamical analysis}\label{de}

We will consider that DE is described by  single real scalar field (quintessence) with energy density $\rho_\phi$ and pressure $p_\phi$, whose equation of state is $w_\phi=p_\phi/\rho_\phi$. DE is interacting with DM through a transfer of energy-momentum between them,  such that that total energy-momentum is conserved. In the flat Friedmann--Lama\^itre--Robertson--Walker (FLRW) background with a scale factor $a$, the continuity equations for both components and for radiation are
\begin{eqnarray}\label{contide}
\dot{\rho}_\phi+3H(\rho_\phi+p_\phi)&=-\mathcal{Q}\,,\nonumber\\
\dot{\rho}_m+3H\rho_m&=\mathcal{Q}\,,\nonumber\\
\dot{\rho}_r+4H\rho_r&=0\,,
\end{eqnarray}
 respectively, where $H=\dot{a}/a$ is the Hubble rate,  $\mathcal{Q}$ is the coupling between DM and DE and the dot is a derivative with respect to the cosmic time.  A positive $\mathcal{Q}$ corresponds to DE being transformed into DM, while negative $\mathcal{Q}$ means the transformation in the opposite direction. In principle, the coupling  can depend upon several variables $\mathcal{Q}=\mathcal{Q}(\rho_m,\rho_\phi,$ $ \dot{\phi},H,t,\dots)$, thus we assume here the first form used in the literature  $\mathcal{Q}=Q \rho_m\dot{\phi}$ \cite{Wetterich:1994bg,Amendola:1999er}, where $Q$ is a positive constant (a negative constant would give similar results). A coupling of the form $Q \rho_\phi\dot{\phi}$ would have no cosmological perturbations, because as we will point out later, DE is expected not to cluster at sub-horizon scales \cite{Duniya:2013eta}. On the other hand, a coupling $Q (\rho_\phi+\rho_m)\dot{\phi}$ was shown not to be viable to describe all three cosmological eras \cite{Landim:2016gpz}.

To deal with the dynamics of the system, we will define dimensionless variables. The new variables are going to characterize a system of differential equations in the form
\begin{equation}
X'=f[X]\,,
\end{equation}
\noindent where $X$ is a column vector of dimensionless variables and the prime is the derivative  with respect to $ \log a$, where we set the present scale factor $a_0$ to be one. The critical points $X_c$ are those ones that satisfy $X'=0$. In order to study stability of the fixed points, we consider linear perturbations $Y$ around them, thus $X=X_c+Y$. At the critical point the perturbations $Y$ satisfy the following equation 
\begin{equation}
Y'=\mathcal{J}Y\,,
\end{equation}
\noindent where $\mathcal{J}$ is the Jacobian matrix. The eigenvalues of $\mathcal{J}$ determine if the critical points are stable (if all eigenvalues are negative), unstable (if all eigenvalues are positive) or saddle points (if at least one eigenvalue is positive and the others are negative, or vice-versa).

\section{Quintessence field dynamics}\label{quintphantom}

The real canonical scalar field $\phi$ is described by the  Lagrangian
\begin{equation}\label{scalar}
 \mathcal{L}=-\sqrt{-g}\left(\frac{1}{2}\partial^\mu\phi\partial_\mu\phi+V(\phi)\right)\,,
\end{equation} 

\noindent where $V(\phi)=V_0 e^{-\lambda \phi}$ is the potential and $V_0$  and $\lambda>0$ are constants. A negative $\lambda$ is obtained if the field is replaced by $\phi\rightarrow -\phi$, thus we may restrict our attention to a positive $\lambda$. For a homogeneous field $\phi\equiv\phi(t)$ in an expanding Universe with FLRW metric and scale factor $a\equiv a(t)$, the equation of motion becomes
\begin{equation}\label{eqmotionscalar1}
 \ddot{\phi}+3 H\dot{\phi}+V'(\phi)=-Q \rho_m\,.
\end{equation} 
In the presence of matter and radiation, the Friedmann equations are
\begin{equation}\label{eq:1stFEmatterS}
  H^2=\frac{1}{3}\left(\frac{\dot{\phi}^2}{2}+V(\phi)+ \rho_m+\rho_r\right)\,,
\end{equation}

\begin{equation}\label{eq:2ndFEmatterS}
  \dot{H}=-\frac{1}{2}\left(\dot{\phi}^2+\rho_m+\frac{4}{3}\rho_r\right)\,,
\end{equation}

\noindent and the equation of state becomes
\begin{equation}\label{eqstateS}
 w_\phi=\frac{p_\phi}{\rho_\phi}=\frac{\dot{\phi}^2-2 V(\phi)}{\dot{\phi}^2+2 V(\phi)}\,.
\end{equation}
Cosmological perturbations are the roots of structure formation, and they are reached perturbing the energy-momentum tensor and the metric. It is convenient to work in the conformal (Newtonian) gauge, where the density perturbation $\delta\equiv \delta\rho/\rho$ and the 
divergence of the velocity perturbation in Fourier space $\theta\equiv a^{-1}i k^j\delta u_j$ obey the following equations for a general interacting DE model \cite{Marcondes:2016reb}
\begin{eqnarray}\label{eq:DMperturbation}
&\dot{\delta}+\Big[3H(c^2_s-w)-\frac{\mathcal{Q}}{\rho} \Big]\delta+(1+w)(\theta-3\dot{\phi})=-\frac{\delta\mathcal{Q}}{\rho}\,,\nonumber\\
&\dot{\theta}+\Big[H(1-3w)-\frac{\mathcal{Q}}{\rho}+\frac{\dot{w}}{1+w} \Big]\theta-k^2\phi-\frac{c_s^2}{1+w}k^2\delta=0\,,
\end{eqnarray}
where $\phi$ is the metric perturbation in Newtonian gauge, $c_s\equiv\delta p/\delta\rho$ is the sound speed and $ k^i$ are the components of the wave-vector in Fourier space. For radiation the density fluctuations do not cluster, and for quintessence $c_s=1$ and DE perturbations are expected to be negligible at sub-horizon scales \cite{Duniya:2013eta}, thus they can be neglected. It is interesting, therefore, to analyse only DM perturbations, and to do so it is more convenient to merge Eq. (\ref{eq:DMperturbation}) into a second-order differential equation. This is done using the Poisson equation
\begin{equation}
    k^2\phi=-\frac{3}{2}H^2\Omega_m\delta_m\,,
\end{equation}
whose result gives
\begin{equation}
    \ddot{\delta}_m+(2H-Q\dot{\phi})\dot{\delta}_m-\frac{3}{2}H^2\Omega_m\delta_m=0\,.
\end{equation}
Now we may proceed to the dynamical analysis of the system.

\section{Autonomous system} \label{auto}

The new dimensionless variables are defined as

\begin{eqnarray}\label{eq:dimensionlessXYS}
 x\equiv  &\frac{\dot{\phi}}{\sqrt{6}H}, \quad y\equiv \frac{\sqrt{V(\phi)}}{\sqrt{3}H}, \quad z\equiv \frac{\sqrt{\rho_r}}{\sqrt{3}H}, \quad \lambda\equiv -\frac{V'}{V},\nonumber\\
  & \quad \Gamma\equiv \frac{VV''}{V'^2},\quad U_m\equiv \frac{\delta'_m}{\delta_m}\,,
\end{eqnarray}
where the prime is the derivative with respect to $N\equiv \ln a$.

The DE density parameter is written in terms of these new variables as
\begin{equation}\label{eq:densityparameterXYS}
 \Omega_\phi \equiv \frac{\rho_\phi}{3H^2} = x^2+y^2\,,
 \end{equation}
thus the first Friedmann equation (\ref{eq:1stFEmatterS})  becomes 
\begin{equation}\label{eq:SomaOmegasS}
\Omega_\phi+\Omega_m+\Omega_r=1\,,
\end{equation}
 where the matter and radiation density parameter are defined by $\Omega_i=\rho_i/(3H^2)$, with $i=m,r$. From Eqs. (\ref{eq:densityparameterXYS}) and (\ref{eq:SomaOmegasS}) $x$ and $y$ are restricted in the phase plane $x^2+y^2\leq 1$.

The equation of state $w_\phi$ is written in terms of the dimensionless variables as
\begin{equation}\label{eq:equationStateXYS}
 w_\phi =\frac{\epsilon x^2+-y^2}{\epsilon x^2+y^2}\,,
\end{equation}
and the total effective equation of state is
\begin{equation}\label{eq:weffS}
 w_{eff} = \frac{p_\phi+p_r}{\rho_\phi+\rho_m+\rho_r}= x^2-y^2+\frac{z^2}{3}\,,
\end{equation}
\noindent with an accelerated expansion for  $w_{eff} < -1/3$.  The dynamical system for the variables  $x$,  $y$, $z$, $\lambda$ and $U_m$ are 
\begin{eqnarray}\label{dynsystemS}\label{eq:dx1/dnS}
\frac{dx}{dN}&=-3x+\frac{\sqrt{6}}{2}y^2\lambda-\frac{\sqrt{6}}{2} Q(1-x^2-y^2-z^2)\nonumber\\ &
-xH^{-1}\frac{dH}{dN}\,,\end{eqnarray}

\begin{equation}\label{eq:dy/dnS}
\frac{dy}{dN}=-\frac{\sqrt{6}}{2}x y\lambda-yH^{-1}\frac{dH}{dN}\,,
\end{equation}

\begin{equation}\label{eq:dz/dnS}
\frac{dz}{dN}=-2z-zH^{-1}\frac{dH}{dN}\,,
\end{equation}

\begin{equation}\label{eq:dlambda/dnS}
\frac{d\lambda}{dN}=-\sqrt{6}\lambda^2 x\left(\Gamma-1\right)\,,
\end{equation}
\begin{eqnarray}\label{eq:dlU/dnS}
\frac{d U_m}{dN}&=-U_m(U_m+2) +\frac{3}{2}(1-x^2-y^2-z^2)\nonumber\\
&+\sqrt{6}QxU_m-U_mH^{-1}\frac{dH}{dN}\,,
\end{eqnarray}
 where
\begin{equation}
H^{-1}\frac{dH}{dN}=-\frac{3}{2}(1+ x^2-y^2)-\frac{z^2}{2}\,.
\end{equation}

\subsection{Critical points}

 The fixed points of the system are obtained by setting $dx/dN=0$, $dy/dN=0$, $dz/dN$, $d\lambda/dN=0$ and $dU_m/dN=0$ in Eqs. (\ref{dynsystemS})--(\ref{eq:dlU/dnS}). When $\Gamma=1$, $\lambda$ is constant the potential is $V(\phi)=V_0e^{-\lambda \phi}$ \cite{copeland1998,ng2001}. The fixed points for coupled \cite{Amendola:1999er} or uncoupled quintessence  \cite{copeland1998} are well-known in the literature,  and only the critical points that may satisfactorily represent  one of the three cosmological eras (radiation-dominated, matter-dominated or DE-dominated) are shown in Table \ref{criticalpointsS} (see \cite{Bahamonde:2017ize} for a review). For those points, the additional critical point $U_m$ was found.

 \begin{table*}\centering
\begin{tabular}{cccccccc}
\hline\noalign{\smallskip}
Point  &  $x$ & $y$  &$z$& $U_m$& $w_\phi$ & $\Omega_\phi$& $w_{eff}$ \\\\
\noalign{\smallskip}\hline\noalign{\smallskip}
		(a1) 	 &$-\frac{\sqrt{6}Q}{3}$   & 0& 0&$-\frac{1}{4}\left(1+2 Q^2+\sqrt{4 Q^4-12 Q^2+25}\right)$ & 1& $\frac{2Q^2}{3}$&  $\frac{2Q^2}{3}$ \\
			(a2) 	 &$-\frac{\sqrt{6}Q}{3}$   & 0& 0&$-\frac{1}{4}\left(1+2 Q^2-\sqrt{4 Q^4-12 Q^2+25}\right)$ & 1& $\frac{2Q^2}{3}$&  $\frac{2Q^2}{3}$ \\
		(b) & $0$&$0$    & 1 &0 &-- &0 & $\frac{1}{3}$ \\      (c1)& $\frac{\sqrt{6}}{2(\lambda+Q)}$    & $\sqrt{\frac{2Q(Q+\lambda)+3}{2(\lambda+Q)^2}}$ & $0$ & $\frac{Q \left(2-\sqrt{\frac{25 \lambda ^2+4 Q^2+20 \lambda  Q-72}{(\lambda +Q)^2}}\right)-\lambda  \left(1+\sqrt{\frac{25 \lambda ^2+4 Q^2+20 \lambda  Q-72}{(\lambda +Q)^2}}\right)}{4 (\lambda +Q)}$ & $-\frac{Q(Q+\lambda)}{Q(Q+\lambda)+3}$& $\frac{Q(Q+\lambda)+3}{(\lambda+Q)^2}$& $-\frac{Q}{\lambda+Q}$\\
	(c2)& $\frac{\sqrt{6}}{2(\lambda+Q)}$    & $\sqrt{\frac{2Q(Q+\lambda)+3}{2(\lambda+Q)^2}}$ & $0$ & $\frac{Q \left(2+\sqrt{\frac{25 \lambda ^2+4 Q^2+20 \lambda  Q-72}{(\lambda +Q)^2}}\right)-\lambda  \left(1-\sqrt{\frac{25 \lambda ^2+4 Q^2+20 \lambda  Q-72}{(\lambda +Q)^2}}\right)}{4 (\lambda +Q)}$ & $-\frac{Q(Q+\lambda)}{Q(Q+\lambda)+3}$& $\frac{Q(Q+\lambda)+3}{(\lambda+Q)^2}$& $-\frac{Q}{\lambda+Q}$\\
 (d1)&   $\frac{ \lambda}{\sqrt{6}}$ &  $\sqrt{1-\frac{ \lambda^2}{6}}$ & 0&0 & $-1+\frac{ \lambda^2}{3}$ &1  & $-1+\frac{ \lambda^2}{3}$\\
 (d2)&   $\frac{ \lambda}{\sqrt{6}}$ &  $\sqrt{1-\frac{ \lambda^2}{6}}$ & 0&$\frac{1}{4} \left(2 \lambda ^2+4 \lambda  Q-8\right)$ & $-1+\frac{ \lambda^2}{3}$ &1  & $-1+\frac{ \lambda^2}{3}$\\

 \noalign{\smallskip}\hline
\end{tabular}
\caption{\label{criticalpointsS} Critical points ($x$, $y$ $z$, $U_m$)   for  quintessence field.   $U_m$ was found only for the viable points that may describe one of the three cosmological eras.  The table shows the correspondent equation of state for DE (\ref{eq:equationStateXYS}), the effective equation of state (\ref{eq:weffS}) and the density parameter for DE (\ref{eq:densityparameterXYS}).}
\end{table*}

The eigenvalues of the Jacobian matrix were found for each fixed point in Table \ref{criticalpointsS} and the results are shown in Table \ref{stabilityS}. The eigenvalues $\mu_{3c}$ and $\mu_{4c}$ are 
\begin{eqnarray}\label{eigenh}
\mu_{3c,4c}&=-\frac{3 \lambda ^2+9 \lambda  Q+6 Q^2}{4 (\lambda +Q)^2}\pm\frac{\sqrt{3}}{4 (\lambda +Q)}\Big[72-21 \lambda ^2-16 \lambda  Q^3 \nonumber\\&-4Q^2 \left(8 \lambda ^2-15\right)-4 \lambda  \left(4 \lambda ^2-9\right) Q\Big]^{1/2}\,.
\end{eqnarray}

\begin{table*}
\centering
\begin{tabular}{cccccc}
\hline\noalign{\smallskip}
  Point & $\mu_1$ & $\mu_2$ & $\mu_3$ &$\mu_4$ & Stability\\
\noalign{\smallskip}\hline\noalign{\smallskip}
  (a1)        & $\frac{1}{2} \left(2 Q^2-3\right)$& $\frac{1}{2} \left(2 Q^2-1\right)$& $\frac{1}{2} \sqrt{4 Q^4-12 Q^2+25}$ &$\frac{1}{2} \left(2 Q^2+2 \lambda  Q+3\right)$ & saddle\\
   (a2)        & $\frac{1}{2} \left(2 Q^2-3\right)$& $\frac{1}{2} \left(2 Q^2-1\right)$& $-\frac{1}{2} \sqrt{4 Q^4-12 Q^2+25}$ &$\frac{1}{2} \left(2 Q^2+2 \lambda  Q+3\right)$ & saddle\\
  (b)    & $2$&$-1$&$1$& 0& unstable\\
  (c1)  & $-\frac{\lambda +4 Q}{2 (\lambda +Q)}$&$\frac{1}{2} \sqrt{\frac{25 \lambda ^2+4 Q^2+20 \lambda  Q-72}{(\lambda +Q)^2}}$  &$\mu_{3c}$&$\mu_{4c}$ & saddle\\ 
   (c2)  & $-\frac{\lambda +4 Q}{2 (\lambda +Q)}$&$-\frac{1}{2} \sqrt{\frac{25 \lambda ^2+4 Q^2+20 \lambda  Q-72}{(\lambda +Q)^2}}$  &$\mu_{3c}$&$\mu_{4c}$ & saddle\\ 
  (d1)   & $\frac{1}{2} \left(\lambda ^2-6\right)$   &   $\frac{1}{2} \left(\lambda ^2-4\right)$  & $\lambda ^2+\lambda  Q-3$ & $\frac{\lambda ^2}{2}+\lambda  Q-2$ & saddle or stable  \\ 
   (d2)   & $\frac{1}{2} \left(\lambda ^2-6\right)$   &   $\frac{1}{2} \left(\lambda ^2-4\right)$  & $\lambda ^2+\lambda  Q-3$ & $-\frac{\lambda ^2}{2}-\lambda  Q+2$ & saddle or stable  \\

     \noalign{\smallskip}\hline
\end{tabular}
\caption{\label{stabilityS} Eigenvalues and stability of the fixed points.}
\end{table*}

 The points (a1) and (a2) are the so-called ``$\phi$-matter-dominated epoch'' ($\phi$MDE) \cite{Amendola:1999er} and they may describe a matter-dominated universe if $\Omega_\phi=2Q^2/3\ll 1$. Thus $\mu_1$ and $\mu_2$ are negative, while $\mu_3$ and $\mu_4$ are always positive. Therefore (a1) and (a2) are saddle points. In the small  $Q$ limit, described above, we have  $U_m=1-\frac{4 Q^2}{5}$ for (a1) and $U_m=-\frac{3}{2}-\frac{Q^2}{5}$ for (a2). The point (a1)  correctly describes the growth of the perturbation $\delta_m\sim a$, with a small correction due to the coupling with DE. On the other hand, the point (a2) does not describe the expected growth of structures.

The radiation-dominated Universe is described by the critical point (b) and  matter perturbations do not increase during this epoch ($U_m=0)$. For this case, the fixed point is unstable, because it has one positive, one negative and one zero eigenvalue \cite{Boehmer:2011tp}. At the background level (without the variable $U_m$) (b) is a saddle point, which indicates that the presence of linear cosmological perturbations drives the point away from the unstable equilibrium (\textit{i.e.} saddle).

\begin{figure}
    \centering
    \includegraphics[scale=0.5]{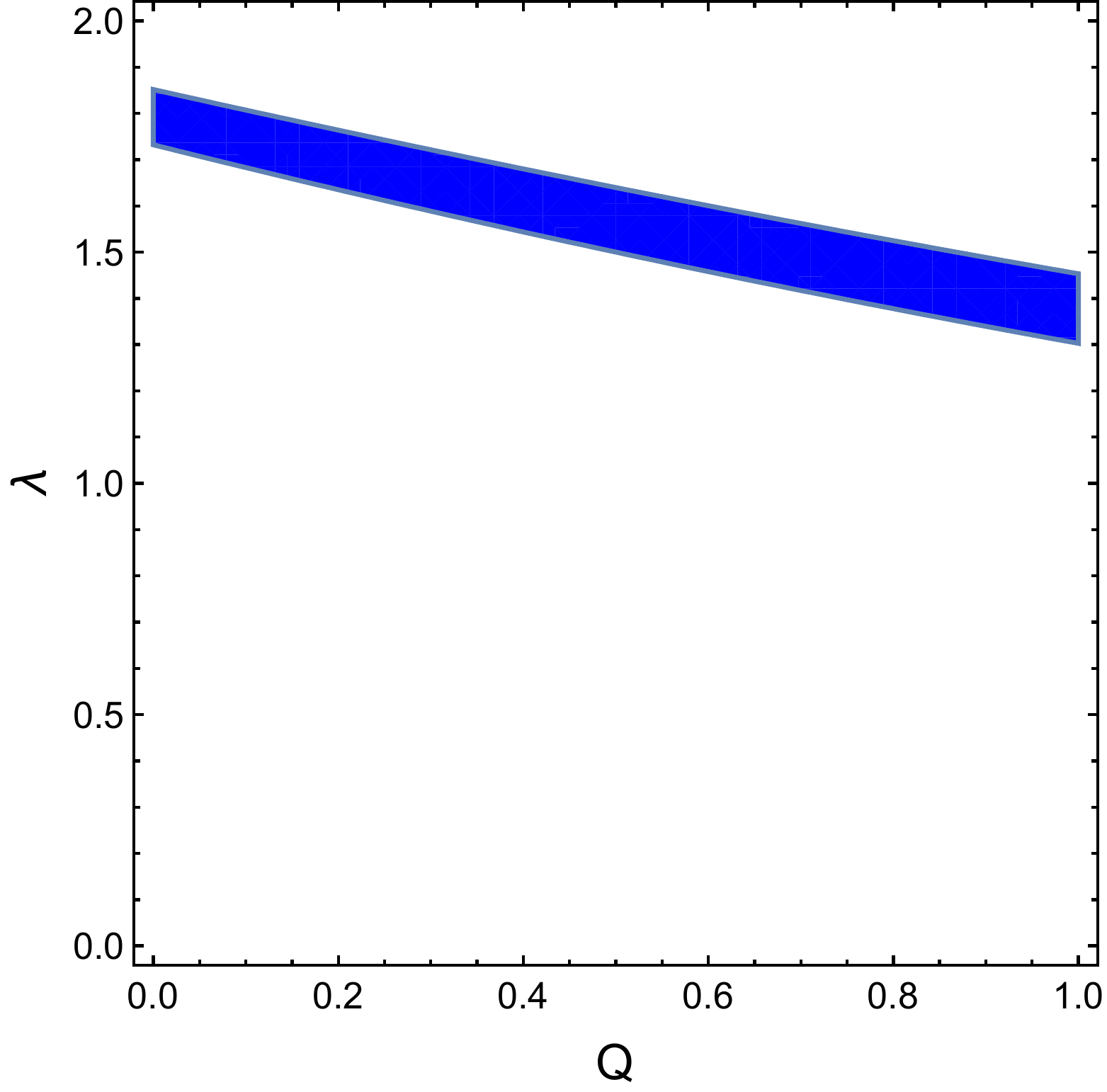}
    \caption{Allowed parameter space for $Q$ and $\lambda$ in order for (c2) to be a stable point. These possible values of $\lambda $ are ruled out by current cosmological observations \cite{Akrami:2018ylq}.}
    \label{fig:plot1}
\end{figure}

At first glance one might think that the points (c1) and (c2) could describe a matter-dominated universe, when $Q\ll \lambda$. In this situation, we would have $U_m=-\frac{1}{4}\left(1\pm\sqrt{25-\frac{72}{\lambda ^2}}\right)$, which is real for $\lambda >\frac{6 \sqrt{2}}{5}$. This value for the parameter  $\lambda$ was excluded by cosmological observations more than one decade ago \cite{Kallosh:2002gf,LopesFranca:2002ek}, being  $\lambda>1$ ruled out by at least  the 3$\sigma$  level \cite{Akrami:2018ylq}. Without cosmological perturbations the fixed points (c)  might represent a DE-dominated Universe, although the match between the coupling constant in this case and  the points (a) is difficult. In our case, both points cannot describe the late accelerated expansion of the Universe. The reason is that (c1) is a saddle point, because $\mu_2$ is always positive while the other eigenvalues can be negative. On the other hand, (c2) could be a stable point if $\mu_4$ were negative. This condition would be satisfied for the set of values $Q$ and $\lambda$ shown in Fig. \ref{fig:plot1}, but as we said before, these values of $\lambda$ are already excluded by current observations. Therefore (c2) is also a saddle point. 

Points (d1) and (d2) exist for $\lambda^2<6$, they can describe the accelerated expansion of the Universe if $\lambda^2<2$ (because $w_{eff}<-1/3$) and they are either saddle or stable, depending on the values of $Q$ and $\lambda$. The point (d1) is an attractor if $Q<\frac{4-\lambda ^2}{2 \lambda }$, while (d2) is attractor for $\frac{4-\lambda ^2}{2 \lambda }<Q<\frac{3-\lambda ^2}{\lambda }$. The matter perturbation is constant ($U_m=0$) for (d1) or decrease for (d2), indicating that the formation of structures does not happen in the DE-dominated Universe.

 \section{Conclusions}\label{conclu}
 
 In this paper we have used dynamical system theory to analyse the evolution of cosmological (matter) perturbations for interacting quintessence. Previous results in the literature \cite{Amendola:1999er}, regarding the possible fixed points that represent one of each cosmological eras, are maintained (with the exception of point (c), which no longer can describe a DE-dominated Universe) and the viable cosmological transition radiation $\rightarrow$ matter $\rightarrow$ DE is achieved considering the sequence of critical points (a1) $\rightarrow$  (b) $\rightarrow$ (d1) or (d2).  The stability of these points remain similar to previous studies, \textit{i.e.}, to those ones considering only background evolution, for both coupled and uncoupled cases. 
 Future constraints on the parameter $\lambda$ will elucidate whether quintessence can still be a DE candidate with a exponential potential or not.

\begin{acknowledgements}
This work was supported by CAPES under the process 88881.162206/2017-01 and Alexander von Humboldt Foundation. \end{acknowledgements}

\bibliographystyle{unsrt}
\bibliography{references}\end{document}